# Resolving Clinicians Queries Across a Grids Infrastructure

F Estrella[1], C del Frate[2], T Hauer[1], R McClatchey[1], M Odeh[1], D Rogulin[1], S R Amendolia[3], D Schottlander[4], T Solomonides[1], R Warren[5]

[1]CCCS Research Centre, University of the West of England, Frenchay, Bristol BS16 1QY, UK
[2]Istituto di Radiologia, Università di Udine, Italy
[3]ETT Division, CERN, 1211 Geneva 23, Switzerland
[4]Mirada Solutions Limited, Mill Street, Oxford, OX2 0JX, UK
[5]Breast Care Unit, Addensbrooke Hospital, Cambridge, UK

**Corresponding author:** Professor Richard H McClatchey, CCCS Director, University of the West of England, Coldharbour Lane, Frenchay, Bristol BS16 1QY, UK
Email: richard.mcclatchey@uwe.ac.uk Tel: +44 117 328 3176, Fax: +44 117 328 2734

**Abstract:** The past decade has witnessed order of magnitude increases in computing power, data storage capacity and network speed, giving birth to applications which may handle large data volumes of increased complexity, distributed over the internet. Medical image analysis is one of the areas for which this unique opportunity likely brings revolutionary advances both for the scientist's research study and the clinician's everyday work. Grids [1] computing promises to resolve many of the difficulties in facilitating medical image analysis to allow radiologists to collaborate without having to co-locate. The EU-funded MammoGrid project [2] aims to investigate the feasibility of developing a Grid-enabled European database of mammograms and provide an information infrastructure which federates multiple mammogram databases. This will enable clinicians to develop new common, collaborative and co-operative approaches to the analysis of mammographic data. This paper focuses on one of the key requirements for large-scale distributed mammogram analysis: resolving queries across a grid-connected federation of images.

## 1. Introduction

Breast cancer as a medical condition, and mammograms as images, are extremely complex with many dimensions of variability across the population. Similarly, the way diagnostic systems are used and maintained by clinicians varies between imaging centres and breast screening programmes, and in consequence so does the appearance of the mammograms generated. It is necessary to understand this variability to be able to study the epidemiology of breast cancer and enhance the usefulness of mammography breast screening by integrating Computer Aided Diagnostic (CAD) tools [3] and quality control [4], [5] in the process. A geographically distributed database that reflects the spread of pathologies across the European population is an essential tool for the epidemiologist and the understanding of the variation in image acquisition protocols is invaluable to the end-user who runs a screening programme.

In order to make the most of such a database it is necessary to have the right tools. This requires an infrastructure to make the large volume of data available to all the centres in an acceptable time, a capable data-mining engine that enables queries based on patient details and text annotations, standardization software to enable the comparison of images from different patients and centres, image analysis algorithms that provide quantitative information, which is otherwise unavailable from visual inspection alone, and detection systems that help in visual diagnosis.

Usually, related personal and clinical information is important (age, gender, selection criteria, disease status). The number of parameters that affect the appearance of an image is so large that the database of images developed at any single site – no matter how large – is unlikely to contain a set of exemplars in response to any given query (e.g. "show me all women in their 50s that developed a tumor within 5 years of starting Hormone Replacement Therapy") that is statistically significant. Overcoming this problem implies constructing a huge, multi-centre – federated – database, while overcoming statistical biases such as lifestyle and diet leads to a database that transcends national boundaries. For ***any*** medical condition, there are potential gains from a pan-national database – so long as that (federated) database is as usable as if it were installed in a single site. Such a database can be suitably supported through a Grid system [1] where many hospital sites are linked together to provide an interconnected collection of patient records and clinical processing power [6] and the collection acts as a single pan-national medical information system.

## 2. MammoGrid User Requirements

The main output of the MammoGrid project, a Grid-enabled software platform (called the MammoGrid Information Infrastructure) which federates multiple mammogram databases, enables clinicians to develop new common, collaborative approaches to the analysis of mammograms. This is being achieved through the use of Grid-compliant services for managing massively distributed files of mammograms, for handling the distributed execution of mammogram analysis software, for the development of Grid-aware algorithms and for the sharing of resources between multiple collaborating medical centres. All this is delivered via a novel software and hardware information infrastructure that guarantees the integrity and security of the medical data.

The MammoGrid project has been driven by the requirements of its user community (represented by Udine (Italy) and Cambridge (UK) hospitals along with medical imaging expertise from Oxford) that have been elicited and specified in detail elsewhere [6]. Rational's Unified Process model (RUP) [7] has been used in the requirements specification for MammoGrid and in particular key requirements engineering activities. The process has identified major use-case scenarios in the use of a distributed database of mammograms deployed across a pan-European Grid and that later can be used to prove the MammoGrid prototype.

The resulting MammoGrid User Requirements Specification (URS) details two essential objectives that must be supported and tested in the MammoGrid project:

- The support of clinical research studies through access to and execution of algorithms on physically large, geographically distributed and potentially heterogeneous sets of (files of) mammographic images, just as if these images were locally resident.
- The controlled and assured access for educational/commercial companies to distributed mammograms for testing novel medical imaging diagnostic technologies in scientifically acceptable clinical trials that fulfill the criteria of evidence-based medical research.

The requirements elicitation process was carried out in consultation with the user community at hospitals in Udine, Cambridge and Torino. In these discussions nine core use cases with corresponding actors have been identified. The use case concerned with mammographic analysis describes the tasks that the 'Mammogram Analyst' actor, normally a radiologist, may undertake to annotate and/or view mammograms and patient details and to execute radiological or epidemiological queries, including use of computer aided detection (CADe) software. This use-case provides the frame for the queries which is the subject of the further sections of this paper.

*Actor:* Mammogram Analyst
*Pre-Conditions*
User-Authentication
*Non-functional requirements:*
CAD Software Interface Requirements
*Flow of Events:*
as per selection of the Mammogram Analyst to link to the appropriate extension point below
*Extension Points:*
(1) View Mammogram and Patient Details
(2) Annotate Mammograms and Patient Details
(2.1) Diagnose Study
(2.2) Diagnose Series
(2.3) Annotate Image
(2.4) Request Computer Aided Detection (CAD)
(2.5) Link Annotations
(2.6) Request CAD in Mammogram Region
(3) Execute Radiological Queries
(3.1) Formulate Radiological Query
(3.2) Refine Radiological Query
*Alternative Flows:*
(1) Unsuccessful User Authentication
(2) CAD Interface Error
(3) Invalid Query Selections
*Post-Conditions*
(1) Mammogram Image Annotated
(2) Patient Details Changed
(3) Results of Query Execution (Grid)
*Use-Case View:*

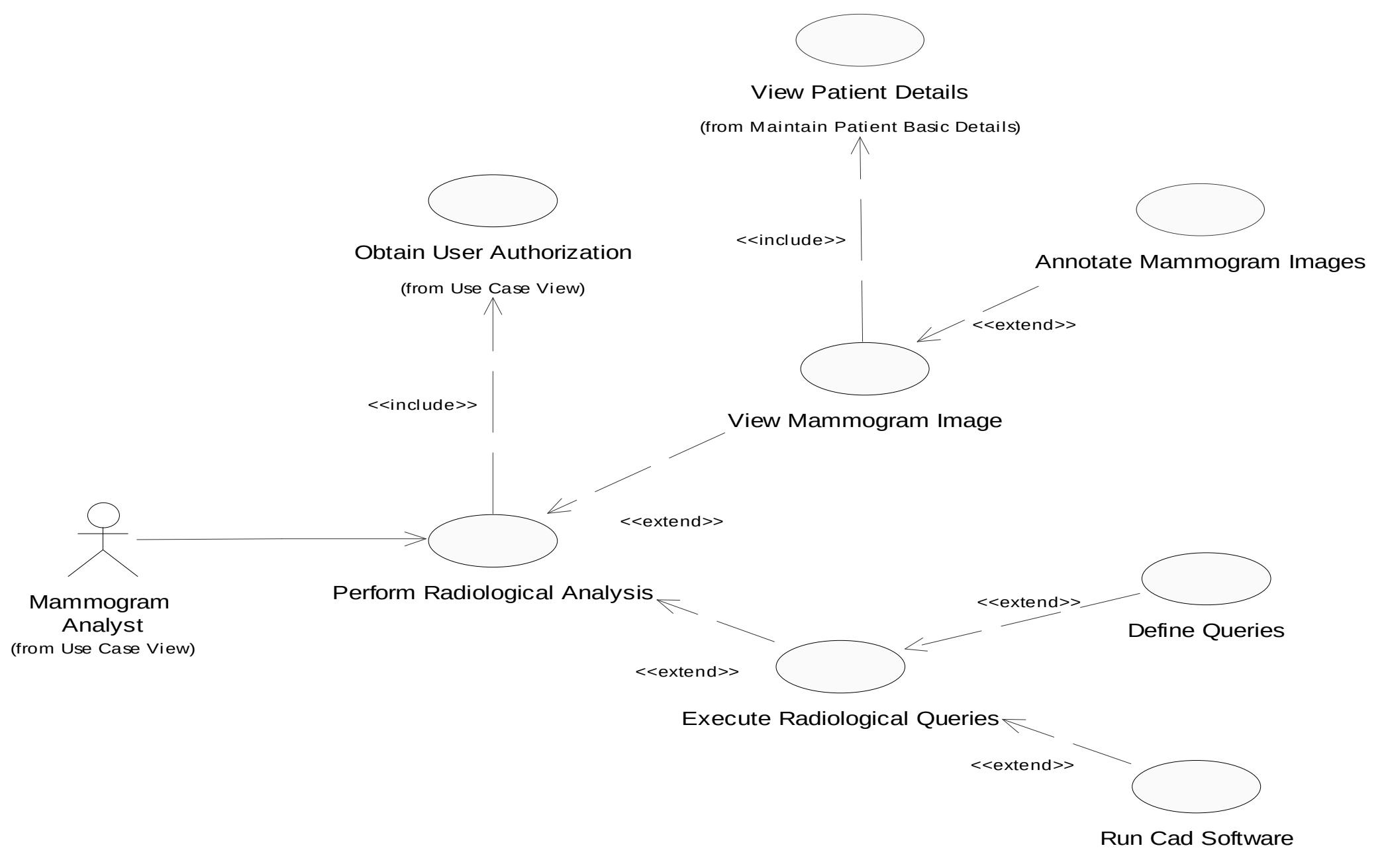

**Figure 1:** The Mammogram Analyst Use Case diagram

## 3. Resolving Clinicians' Queries

In the MammoGrid proof-of-concept demonstrator, clinician queries will be handled and resolved against data resident in a Grids infrastructure. User Requirements have been gathered that will enable queries to be executed and data retrieved for the analysis of mammograms. In particular the MammoGrid project will test the access to sets of mammogram images for the purposes of breast density assessment and for the testing of CADe studies of mammograms.

Queries can be categorized into simple and complex queries. Simple queries use predicates that refer to simple attributes of meta-data saved alongside the mammographic images. One example of a simple query might be to 'find all mammograms for women aged between 50 and 55' or 'find all mammograms for all women over 50 undergoing HRT treatment'. Provided that age and HRT related data is stored for (at least a subset of) patients in the patient meta-data then it is relatively simple to select the candidate images from the complete set of images either in one location of across multiple locations. It is also possible to collect data concerning availability of requested items so as to inform the design of future protocols, thus engineering a built-in enhancement process.

There are, however, queries which refer to data that has not been stored as simple attributes in the meta-data but rather require derived data to be interrogated or an algorithm to be executed. Examples of these might be queries that refer to the semi-structured data stored with the images through annotation or clinician diagnosis or that is returned by, for example, the execution of the CADe image algorithms.

### *3.1. Typical Complex MammoGrid Queries*

This section describes three example use-cases illustrating the nature of complex queries which the MammoGrid infrastructure should handle.

#### 3.1.1. Use Case 1: Patient's first visit

Consider a patient on her first visit to the mammography center (following referral by GP, worried about a symptom). The typical workflow of the visit looks like this:

- Mammograms (2 ´ MLO and 2 ´ CC[1]) taken
- Radiologist reads them and annotates[2] left MLO (LMLO) and left CC (LCC)
- Radiologist requests CADe for LMLO and LCC images.
- *Query:* Radiologist requests 'find similar cases'. Example criteria might include women:
    - of same age ± 3
    - with same number of children (0), (1-2), (3-4), (5+)
    - with same age ranges of children (equivalently, age at first and last pregnancy)
    - with images that the algorithm "find one like it" matches well either in MLO or CC
- Radiologist reviews demographics and personal data and determines best four cases to request images.
- Radiologist reviews comparable images with histories and analyses:
    - consider the best match
    - take images from first diagnosis to current state
    - review growth of lesion (ideally identifies the lesion across images)

#### 3.1.2. Use Case 2: Epidemiology Study

Consider an epidemiologist who is conducting a study on contralateral breast cancer. The typical queries she is interested in running may include:

---

[1] MLO – Medio-Lateral Oblique, taken at 45º from shoulder to opposite hip; CC – Cranio-Caudal, taken vertically down from above.

[2] Annotation – a region is marked out as suspect or for further analysis.

- Find all patients in the distributed database who have developed cancer in the other breast after successful therapy (specific or otherwise) on the first cancer.
- Consider mammographic features from the time of first diagnosis and any correlation to occurrence of contralateral cancer.
- Consider measures of asymmetry and their correlation to contralateral cancer.

#### 3.1.3. Use Case 3: Quality Control of Radiology Diagnosis

Consider the use case of comparative study of radiologists' annotations. The typical queries which can be used to survey radiologists' diagnostic processes include the following example queries:

- For a period of six months, allocate each patient who attends for screening at random to two out of three radiologists so that all three possible pairs get roughly equal numbers.
- For each patient, ask both radiologists to examine the mammograms and to make any necessary annotations.
- Submit all annotations for CADe and measure differences between radiologists' annotations and CADe (could be area if masses, counts if microcalcifications) and between the two radiologists in each case.
- Consider correlation to experience, the length of the viewing session and the serial order of the given image in that session, and the radiologist's perception whether this was the first or second reading.

### *3.2. The Role of Meta-Data*

During the final phase of implementation and testing, lasting until the completion of the project, the meta-data structures required to resolve the clinicians' queries will be delivered using the meta-modelling concepts of the CRISTAL project [8]. This will involve customizing a set of structures that will describe mammograms, their related medical annotations and the queries that can be issued against these data. The meta-data structures will be stored in a database at each node in the MammoGrid (e.g. at each hospital or medical centre) and will provide information on the content and usage of (sets of) mammograms.

The query handling tool will locally capture the elements of a clinician's query and will issue a query, using appropriate Grids software, against the meta-data structures held in the distributed hospitals. At each location the queries will be resolved against the meta-data and the constituent sub-queries will be remotely executed against the mammogram databases. The selected set of matching mammograms will then be either analyzed remotely or will be replicated back to the centre at which the clinician issued the query for subsequent local analysis, depending on the philosophy adopted in the underlying Grids software. All data objects will reside in standard commercial databases, which will also hold descriptions of the data items.

### *3.3. The Query Handler*

The user will submit queries that are serviced locally and farmed out to available resources when data from the network is required. In resolving queries the system will consult the knowledge it has acquired from previous queries. Data will be immediately returned to the user and the knowledge base updated. New data is processed only when necessary. With this approach the computation required by a domain-specific application is analyzed and farmed out to appropriate data sources rather than moving or replicating potentially vast amounts of data and processing.

The querying software largely constitutes:

Query Manager components

- Query Translator,
- Query Analyser,
- Local Query Handler,
- Remote Query Handler and
- Result Handler

Data Sources

- User's Terms and Mammogrid specific meta-data,
- Local database and
- Stored query database

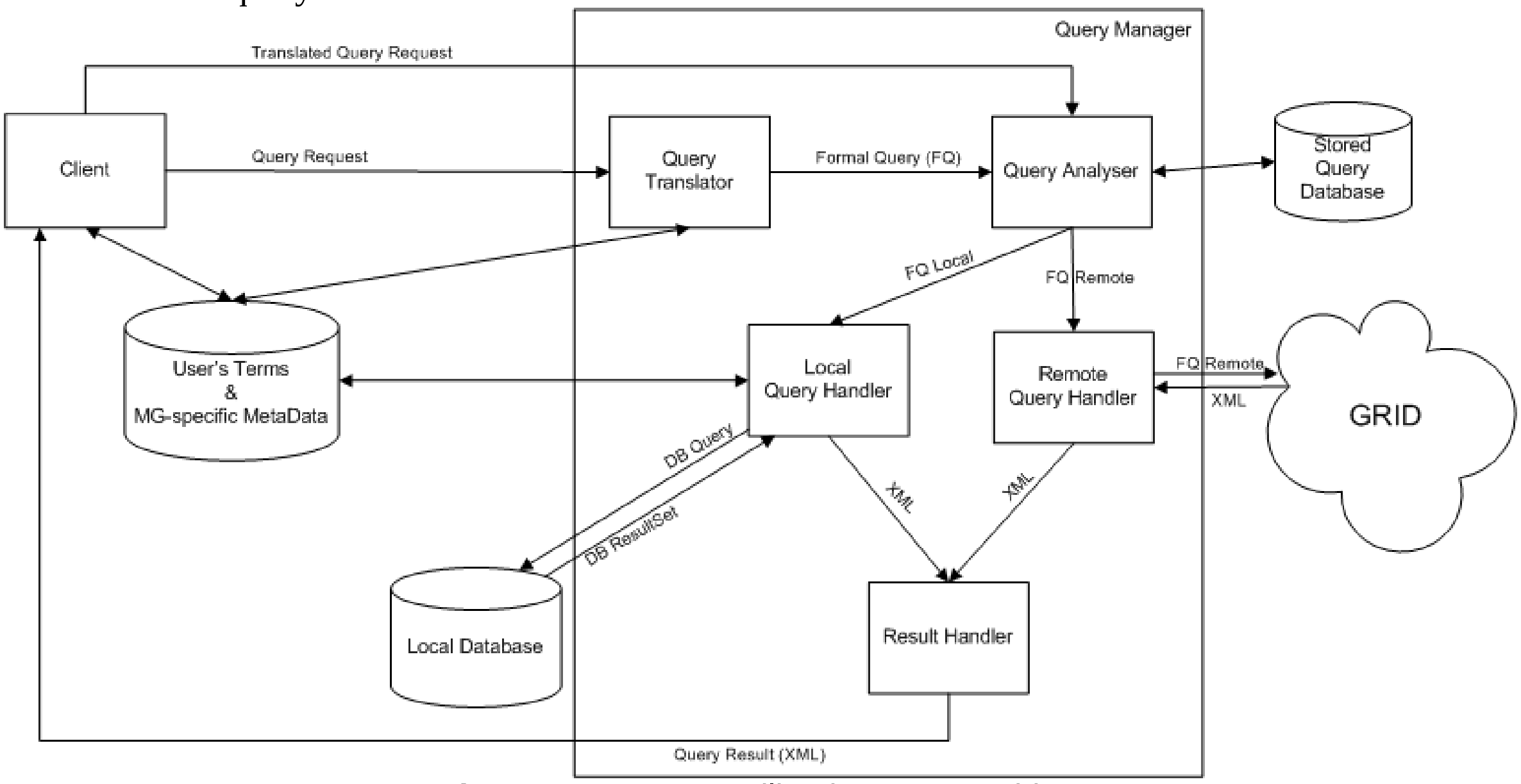

**Figure 2:** Query Handling in Mammogrid

Figure 2 illustrates the query handling and execution in Mammogrid. The sequence of events is as follows:

(1) Clients (e.g. end-users, applications) define their mammogram analysis in terms of queries they wish to be resolved across the collection of data repositories (either locally- or remotely-held data). This uses descriptive information (User's Terms and MG-specific metadata) about the query domain (both graphical specifications and user-specific terms) to translate the user query into a 'data request' using standard terms.

(2) Query Translator takes the user request and translates to a MG-defined formal query representation.

(3) Queries are executed at the location where the relevant data resides. That is, the sub-queries are moved to the data, rather than large quantities of data being moved to the clinician, which is prohibitively expensive given the quantities of data. The Query Analyser takes a formal query representation and de-composes into (a) formal query for local processing and (b) formal query for remote processing. It then forwards these de-composed queries to the Local Query Handler and the Remote Query Handler for the resolution of the request.

(4) The Local Query Handler generates query language statements (e.g. SQL) in the query language of the associated Local DB (e.g. MySQL). The result set is converted to XML and routed to the Result Handler.

(5) The Remote Query Handler is a portal for propagating a queries and results between sites. This handler forwards the formal query for remote processing (3b above) to the Query Analyser of the remote site. The remote query result set is converted to XML and routed to the Result Handler.

(6) The Result Handler is responsible for collecting query results – both local and remote. The query handlers return XML results, and these are “joined” to create the overall result to be sent back to the requestor – either the client of the Remote Query Handler.

Figure 3 shows the propagation of queries between sites.

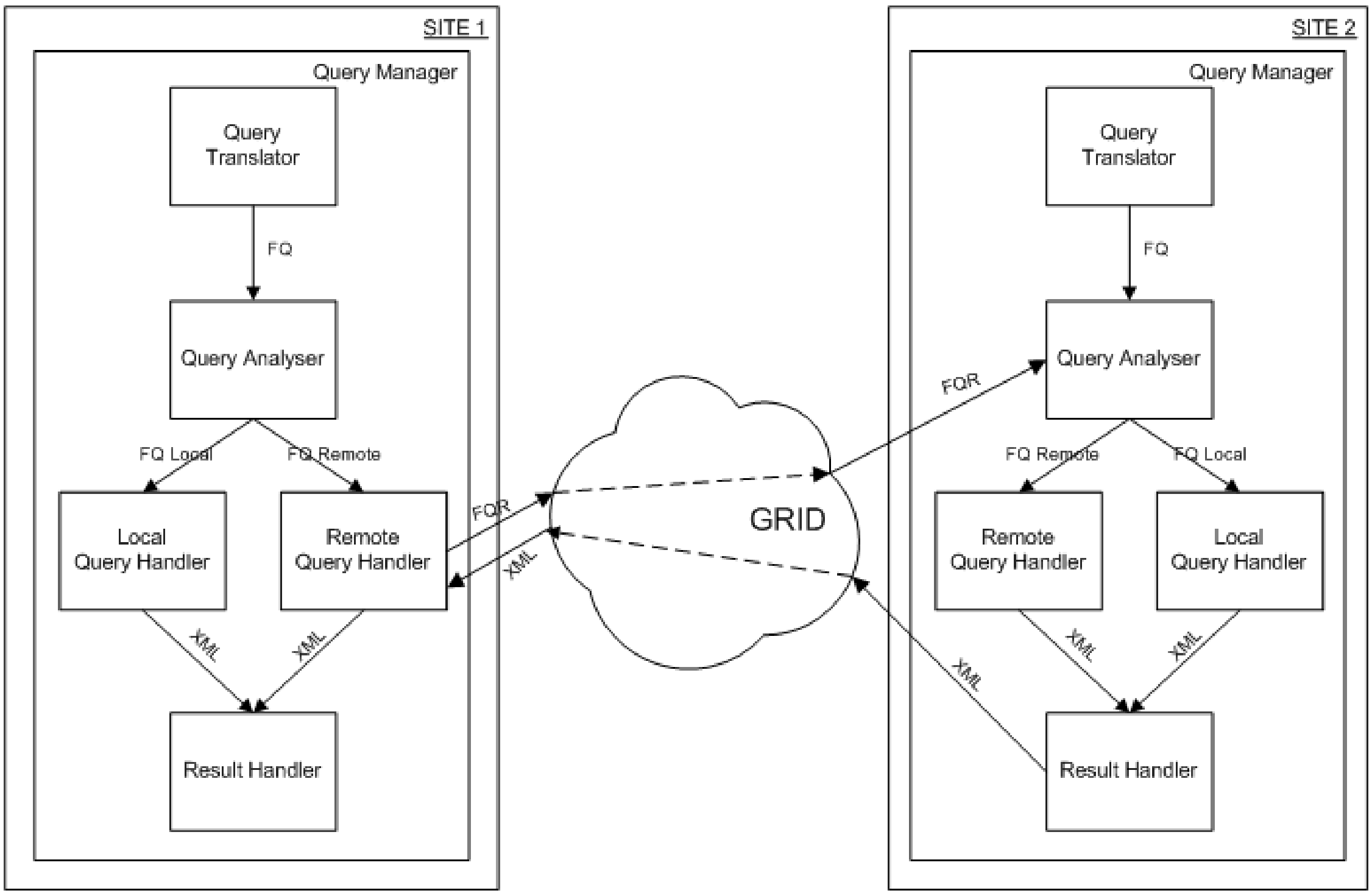

**Figure 3:** Propagation of Queries between Sites

## 4. Current Status and Conclusions

The MammoGrid project has recently delivered its first proof-of-concept prototype enabling clinicians to store (digital) mammograms along with appropriately anonymised patient meta-data and to provide controlled access to mammograms both locally and remotely stored. A typical database comprising several hundred mammograms is being created for user tests of the query handler. The prototype comprises a high-quality clinician visualization workstation [9] used for data acquistion and inspection, a DICOM-compliant interface to a set of medical services [10] (annotation, security, image analysis, data storage and querying services) residing on a so-called ‘Grid-box’ and secure access to a network of other Grid-boxes connected through Grids middleware. Clinicians are being closely involved with these tests and it is intended that a subset of the clinician queries listed in section 3 will be executed to solicit user feedback. Within the next year a rigorous evaluation of the prototype will then indicate the usefulness of the Grid as a platform for distributed mammogram analysis and in particular for resolving clinicans’ queries. The system will be tuned for performance and for security prior to the release of a second prototype at the end of the project in mid 2005. It is intended that the MammoGrid medical services for this second prototype will adhere to emerging Grids standards e.g [11], [12].

The proliferation of information technology in medical sciences will undoubtedly continue, addressing clinical demands and providing increasing functionality. The MammoGrid project is advancing deep inside this territory and exploring the requirements of evidence-based, computation-aided radiology, as specified by medical scientists and practicing clinicians. This paper has emphasized two aspects which are likely to prove essential to the success of such a project: the importance of extensive requirements analysis and a design which caters for the complexity of the data. The very nature of a project like MammoGrid implies that it is inconceivable to define an exhaustive list or even complete classification of all possible queries which the radiologists may need to run against the distributed database. Inevitably, when the user community starts using such a system, the requirements will undergo adjustments and extension. This paper has illustrated the kind of complexity of the expected queries, based on an initial consultation of radiologists. It is proposed that the design followed, with extensive use of meta-data, is both capable of handling such complex queries in an efficient way and flexible enough to adapt to changing requirements. A design which handles queries using a reflexive data model has been presented as the proposed query model for the MammoGrid infrastructure.

In its first year, the MammoGrid project has faced interesting challenges originating from the interplay between medical and computer sciences and has witnessed the excitement of the user community whose expectations from the a new paradigm are understandably high. As the MammoGrid project moves into its final implementation and testing phase, further challenges are anticipated which will test these ideas to the full.